\documentclass{osa-article}
\journal{osajournal}
\articletype{Research Article}

\usepackage[]{lipsum}
\usepackage{siunitx}
\usepackage[labelsep = period]{caption}
\usepackage{subcaption}
\usepackage{footnote}
\usepackage{pgfplots}
\usepackage{floatrow}
\usepackage{float}
\floatstyle{plaintop}
\restylefloat{table}
\newcommand\addplotgraphicsnatural[2][]{%
    \begingroup
    \pgfqkeys{/pgfplots/plot graphics}{#1}%
    \setbox0=\hbox{\includegraphics{#2}}%
    \pgfmathsetmacro{\xfactor}{\wd0/(\pgfkeysvalueof{/pgfplots/plot graphics/xmax} - \pgfkeysvalueof{/pgfplots/plot graphics/xmin})}%
    \pgfmathsetmacro{\yfactor}{\ht0/(\pgfkeysvalueof{/pgfplots/plot graphics/ymax} - \pgfkeysvalueof{/pgfplots/plot graphics/ymin})}\yfactor%
    \pgfmathsetmacro{\xunit}{\xfactor<\yfactor ? 1 : \xfactor/\yfactor}
    \pgfmathsetmacro{\yunit}{\xfactor<\yfactor ? \yfactor/\xfactor : 1}
    \xdef\marshal{%
        \noexpand\pgfplotsset{unit vector ratio*={\xunit\space \yunit}}%
    }%
    \endgroup
    \marshal
    \addplot graphics[#1] {#2};
}   
\pgfplotsset{compat=1.14}
\begin{document}
\title{Microscope objective for imaging atomic strontium with 0.63 micrometer resolution}
\author{I.H.A.~Knottnerus,\authormark{1} S.~Pyatchenkov,\authormark{1} O.~Onishchenko,\authormark{2} A.~Urech,\authormark{1} F.~Schreck,\authormark{1} and G.A.~Siviloglou\authormark{1,3,*}} 
\address{\authormark{1} Van der Waals-Zeeman Institute, Institute of Physics, Faculty of Science, University of Amsterdam, Science Park 904, 1098 XH Amsterdam, The Netherlands \\ \authormark{2} LaserLaB, Department of Physics and Astronomy, VU University, De Boelelaan 1081, 1081 HV Amsterdam, The Netherlands \\ \authormark{3} Guangdong Provincial Key Laboratory of Quantum Science and Engineering, Shenzhen Institute for Quantum Science and Engineering, and Department of Physics, Southern University of Science and Technology, Shenzhen 518055, Guangdong, China}
\email{\authormark{*}microscope461@strontiumBEC.com}

\begin{abstract}
Imaging and manipulating individual atoms with submicrometer separation can be instrumental for quantum simulation of condensed matter Hamiltonians and quantum computation with neutral atoms. Here we present an open-source design of a microscope objective for atomic strontium consisting solely of off-the-shelf lenses that is diffraction-limited for \SI{461}{\nano\meter} light. A prototype built with a simple stacking design is measured to have a resolution of \SI{0.63(4)}{\micro\meter}, which is in agreement with the predicted value. This performance, together with the near diffraction-limited performance for \SI{532}{\nano\meter} light makes this design useful for both quantum gas microscopes and optical tweezer experiments with strontium. Our microscope can easily be adapted to experiments with other atomic species such as erbium, ytterbium, and dysprosium, as with rubidium Rydberg atoms.
\end{abstract}

\section{Introduction}

In the past decade, arrays of ultracold atoms have been established as an excellent system for quantum simulation \cite{Bloch2012QSI} and a promising platform for quantum computation \cite{10.1093/nsr/nwy088,Weiss2017QC}. Imaging and addressing single atoms are essential for progress in this direction and microscopy techniques have been developed for the case of ultracold quantum particles. More specifically, a top-down approach used by quantum gas microscopes \cite{kuhr2016quantum,Bakr2010QGM, Sherson2010QGM} is the loading of already quantum degenerate gases in site-resolved optical lattices, while a bottom-up approach is based on individually filled and controlled optical tweezers \cite{Ashkin1986OL, Bernien2017, Barredo1021, Schlosser2001Sub,Barredo2018ThreeD, Mello2019PRL}. The core necessity for both approaches is the availability of a high-resolution objective that can resolve individual atoms at the single trap level.

The requirements on the performance of such an objective are highly demanding: a long working distance, correction for the typically thick viewport of the vacuum chamber, and high resolution for a set of desired wavelengths, which can range from ultraviolet to infrared. Objectives fulfilling all these requirements are in most cases not found off-the-shelf. Fortunately, in the study of quantum gases, only a single or just a few wavelengths are of interest for atom imaging and manipulation. This is a huge simplification compared to the design of an achromatic objective and it makes home-built objectives a cost-effective option. Our objective is designed for detecting strontium on the ${}^1S_0 \leftrightarrow {}^1P_1$ transition at the wavelength of $\SI{461}{\nano\meter}$ and it should be suitable without modification for other atomic species with nearby transitions, such as ytterbium (\SI{399}{\nano\meter}), erbium (\SI{401}{\nano\meter}), and dysprosium (\SI{421}{\nano\meter}). Recent openly available designs attempted to exploit this characteristic of atomic systems, but their focus was on wavelengths longer than \SI{550}{\nano\meter} \cite{Bennie2013Obj,Li2018Mic} that cannot readily serve the purpose of imaging atomic strontium. 

The alkaline-earth element strontium \cite{Stellmer2009bec} has two-valence electrons and properties that make it suitable for ultraprecise atomic clocks and quantum simulation \cite{Campbell2017DFG,Martin2013Spin}. As was shown in very recent experiments with ultracold strontium in optical tweezers \cite{Cooper2018PRX, Norcia2018PRX, Covey2018ARX, Jackson2019ARX}, and will be demonstrated in the next sections, the $\SI{461}{\nano\meter}$ transition can lead to submicrometer resolution (using the Rayleigh criterion \cite{Saleh2013FoP}) for a numerical aperture that is not particularly high. Single-site imaging for the great majority of ultracold atom experiments means a resolution that can enable imaging of atoms held by the optical lattice potential created by two counter propagating (usually \SI{1064}{\nano\meter}) beams. Image processing can be used to enhance the resolving power of our microscope to obtain single-site detection \cite{Sherson2010QGM}. To be able to simulate physics on the lattice, the objective needs to have a large field of view (FOV) to resolve hundreds of lattice sites and preferably to have a resolution high enough for the manipulation and spin preparation of individual atoms using other relevant wavelengths. For strontium atoms, especially the long-lived ${}^1S_0 \leftrightarrow {}^3P_0$ ($\lambda = \SI{698}{\nano\meter}$) and ${}^1S_0 \leftrightarrow {}^3P_2$ ($\lambda = \SI{671}{\nano\meter}$) transitions are interesting for the purpose of state preparation and manipulation\cite{Ludlow2015oac, Onishchenko2018ARX}. Techniques involving, for example, magnetic field gradients or AC Stark shifts, can permit that even resolutions of around \SI{1}{\micro\meter} are sufficient to already give single-site addressing (removing or changing an atom's spin state without affecting the immediate neighbors) of atoms \cite{Shibata2014osi}. It is also desirable that the depth of field (DOF) is small enough that adjacent layers end up out of focus and do not distort the measurements. Such blurring effects can be minimized by utilizing atom removal techniques similar to the ones used by \cite{Folling2006PRL}. In addition, to avoid over-complicating our apparatus our objective has to be placed outside the ultrahigh vacuum chamber and thus be corrected for the \SI{3.125}{\milli\meter} thick high purity fused silica (Corning HPFS) viewport. This sets a stringent limitation on the proximity of the optics to the atoms and thus a sufficiently large working distance ($>\SI{15}{\milli\meter}$) is required.

 Here, we report on a high-resolution, long working distance, in-house manufactured, multilens microscope objective that can be used to image atomic strontium with the design procedure and the precise specifications openly accessible. Our objective consists of a plastic tube that contains the lenses separated by spacers. The objective is tested by imaging an illuminated test target containing \SI{200}{\nano\meter} diameter pinholes. The resolution of the objective is measured to be \SI{0.63(4)}{\micro\meter} for the primary wavelength of interest (\SI{461}{\nano\meter}) and always lower than \SI{1.1}{\micro\meter} for different relevant colors of the illuminating light. The prototype has a diffraction-limited resolution for wavelengths \SIrange{460}{530}{\nano\meter}. Also, the FOV and the DOF of the objective are explored.

\section{Design of the objective}
\subsection{Optical design} \label{subsec:opticaldesign}
Our design takes \cite{Bennie2013Obj} as a starting point, which in turn was based on the earlier work by Alt \cite{Alt2002Obj}. The increased resolution is a prerequisite for imaging atoms held in optical lattices. The short wavelength used in our case (\SI{461}{\nano\meter}) is advantageous since it makes a moderate-NA objective capable of achieving a resolution high enough for the commonly used lattice spacing of \SI{532}{\nano\meter}. 

The optical design of our objective is performed with commercial software. We summarize here the guiding principles of our approach. While minimizing on the root-mean-square (RMS) spot size and wavefront deviations, the distances between the lenses are varied in two steps: First, every position is taken as a free parameter in the fitting process in order to obtain an estimate for good values, which are then used as starting values for the next optimization steps. Then, the distance between any two lenses is varied individually, while compensating by refocusing the objective (i.e. varying the distance between the first lens and the viewport). The resulting configuration is presented in Table \ref{tab:lensdesign} and a cross section of the objective is shown in Fig. \ref{fig:stackingmodel}. This design is predicted to be diffraction-limited (Strehl ratio larger than 0.8 at focus) for $\lambda=\SI{461}{\nano\meter}$ and to have a numerical aperture $\text{NA}=\SI{0.44}{}$. Using the Rayleigh criterion we obtain a resolution of \SI{639}{\nano\meter} \cite{Saleh2013FoP}. 

In addition to the main feature of submicrometer resolution, the DOF, FOV and tolerances on the placement of the individual lenses are explored, since they are of importance when several layers of atoms separated by less than a micron have to be imaged. The DOF is modeled by translating the objective with respect of to the viewport while keeping the target plane fixed and measuring the resolution of the microscope. The objective has a resolution of around \SI{0.63}{\micro\meter} over a range of around \SI{3.0}{\micro\meter}. This makes expensive nanopositioners unnecessary for our purposes and also does not limit our single-layer resolving ability, since the out-of-focus layers of atoms can in principle be isolated or removed \cite{Shibata2014osi, Folling2006PRL}. Similarly, the FOV is modeled by sweeping over the acceptable range of incidence angles of the incoming light, yielding a maximum FOV of about \SI{200}{\micro\meter}. For a quantum gas microscope with a lattice constant of \SI{532}{\nano\meter} this corresponds to roughly 380 sites. In comparison, recent state-of-the-art experiments with ultracold gases used about 100 sites \cite{Mazurenko2017Anti}.

To determine the tolerances, each relevant parameter (distance, decentering, tilt etc.) is varied for one lens at a time. The objective is finally refocused by adjusting the distance between the first lens and the viewport. The main figure of merit is the RMS spot size of the system. We note that the first lens is not included in the tolerance testing process as the blurring due to the displacement of that lens can be corrected by adjusting the objective as a whole. 

The design is remarkably robust to axial displacements, as all the lenses can be individually displaced more than \SI{150}{\micro\meter} before the RMS spot size exceeds the diffraction limit. Two tolerances that should not be neglected when using commercial lenses is the center thickness and the decenter of the lenses. All lenses have a thickness tolerance of \SI{100}{\micro\meter} and a decenter tolerance of \SI{3}{arcmin}. The decenter is modeled as the radial displacement from the optical axis and the maximum allowed value of \SI{3}{arcmin} is detrimental for the performance, resulting in a RMS spot size exceeding the diffraction limit for almost every lens. If each lens is rotated such that the decenter is pointed in the same direction, the relative decenter decreases and the impact on the performance becomes manageable. This realization guided us to use slits in the objective tube for the adjustment (tilts, decentering, etc.) of the lenses after the assembly of the objective, see Sec. \ref{subsec:mechanicaldesign}.  The numerical checks of the tolerances on the individual lenses are used as requirements for the housing. Furthermore Monte-Carlo simulations show that assembling the microscope objective with arbitrarily rotated lenses leads to a performance within specifications in several ten percent of all cases. That meant that a few trials in placing the lenses would be enough to achieve diffraction-limited imaging as was verified experimentally.

\begin{table}[t]

  \renewcommand{\arraystretch}{1.2}
    \centering
     
    \begin{tabular}{c c c c c c}
        \hline \hline
        \# & Lens Type & Lens name & Manufacturer & \O \:(mm) & Distance (mm)   \\ \hline
        1 & Pos. Meniscus & LAM-459  & Melles Griot & 18.0 & 6.301 \\
        2 & Pos. Meniscus & LE1076 & Thorlabs & 50.8 & 16.094  \\
        3 & Pos. Meniscus & LE1418 & Thorlabs & 50.8  & 33.259 \\
        4 & Biconvex & KBX-151 & Newport & 50.8 & 44.284  \\
        5 & Plano-concave & LC1315 & Thorlabs & 50.8  & 73.108 \\
        \hline \hline
    \end{tabular}
  \caption{The lenses used to create the objective. All these lenses are commercial lenses with anti-reflection coating in the range of \SIrange{350}{700}{\nano\meter}. The last column is the shortest distance of each lens to the air-side facet of the viewport. }\label{tab:lensdesign}
\end{table}
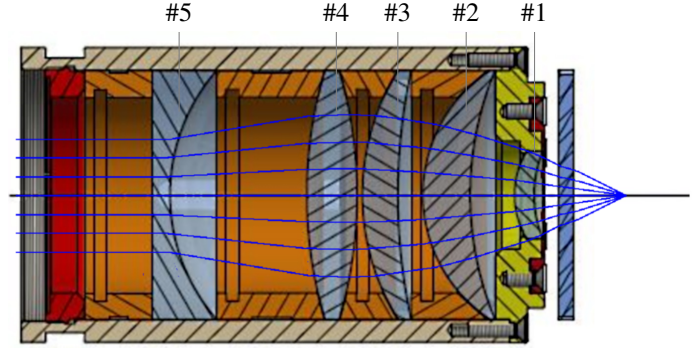
\begin{figure}
  \centering
  \begin{tikzpicture}
    \begin{axis} [
        enlargelimits=false,
        axis equal image,
        axis on top,
        axis lines = none,
        xmin = 0, xmax = 1,
        ymin = 0, ymax = 1,
        width = 0.8\textwidth,
    ]
        \addplotgraphicsnatural[xmin = 0, xmax = 1,
        ymin = 0, ymax = 0.85] {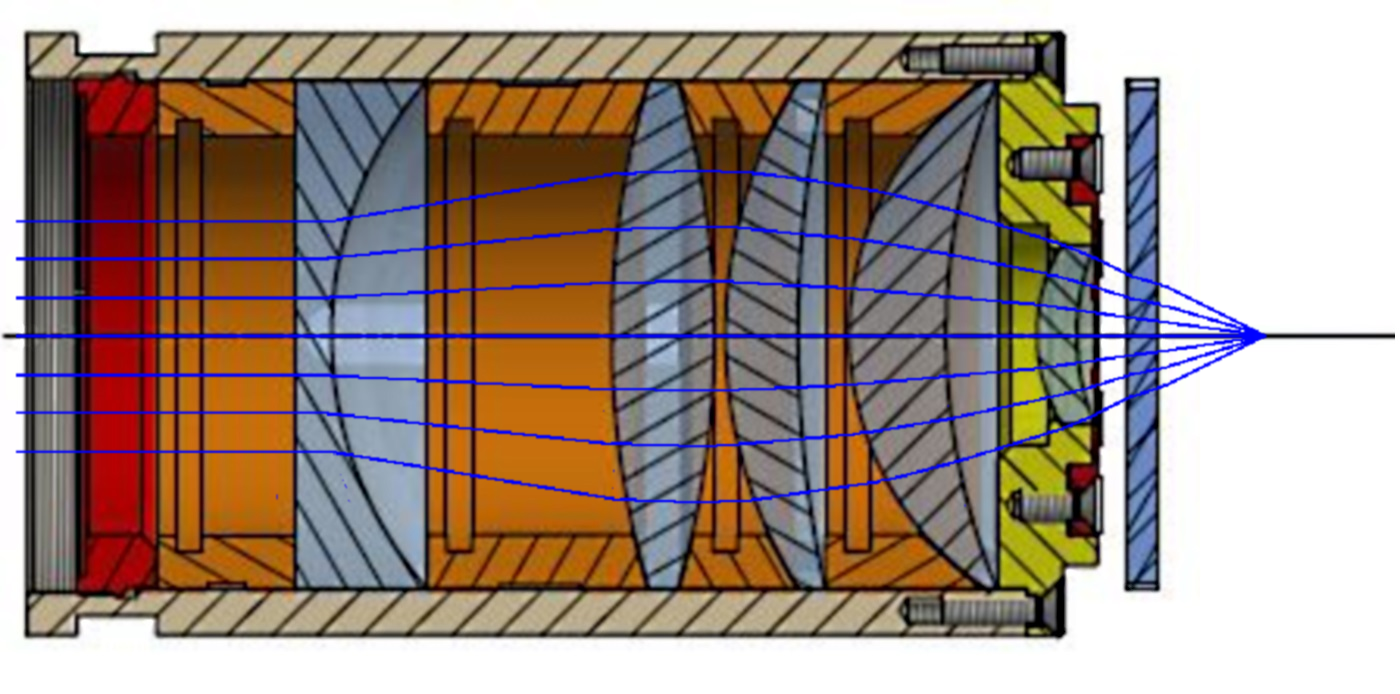};
\draw[gray] (axis cs:0.25,0.85) -- (axis cs:0.25,0.65);
\draw[gray] (axis cs:0.48,0.85) -- (axis cs:0.48,0.65);
\draw[gray] (axis cs:0.57,0.85) -- (axis cs:0.57,0.65);
\draw[gray] (axis cs:0.67,0.85) -- (axis cs:0.67,0.65);
\draw[gray] (axis cs:0.77,0.85) -- (axis cs:0.77,0.55);
        
\draw[color=black](axis cs:0.25,0.9) node[rotate=0] {\#5};
\draw[color=black](axis cs:0.48,0.9) node[rotate=0] {\#4};
\draw[color=black](axis cs:0.57,0.9) node[rotate=0] {\#3};
\draw[color=black](axis cs:0.67,0.9) node[rotate=0] {\#2};
\draw[color=black](axis cs:0.77,0.9) node[rotate=0] {\#1};
  \end{axis}
  \end{tikzpicture}  

  \caption{The stacking design of the objective with modeled rays. Each lens is held in place with high precision spacing rings (orange). The assembly of lenses and spacing rings is fixed inside the tube with a locking ring (red). The first lens is mounted on a separate piece (yellow). The glass plate on the right depicts the viewport. All pieces are made out of PEEK.}
  \label{fig:stackingmodel}

\end{figure}

\subsection{Mechanical design}
\label{subsec:mechanicaldesign}
The optical design targets dictate the requirements on the mechanical construction of the objective. The axial placement of the lenses can be off without detrimental effects by several tens of micrometers and the lens tilt has to be preferably below \SI{0.02}{\degree}. The housing material of the objective has to be selected judiciously by taking into account the in-house manufacturing capabilities and requirements beyond lens placement, for example originating from the needs of atomic physics experiments. To prevent eddy currents from flowing when ramping the electromagnetic coils that surround the objective in our setup, the housing has to be made from a material with very low conductivity. At the same time, the stiffness and the low thermal expansion coefficient of the material are important for the robustness of the optical alignment.

For the housing of the objective a stacking model is chosen, following the example of \cite{Alt2002Obj,Bennie2013Obj} but introducing several additional features as shown in Fig. \ref{fig:stackingmodel}. The lenses are placed inside an outer tube and are held in place by spacing rings that fit precisely between two adjacent lenses. The edges of the spacing rings are linear and at an angle matching the outer shape of the lenses. The whole assembly is locked inside the outer tube by a threaded locking ring at the end. The first lens has a different diameter and is mounted on a separate piece that also serves as the cap of the outer tube. The outer tube is equipped with slits that permit direct access to the lenses and allow us to guide and rotate the lenses during post-assembly adjustments. Six mirrors can be placed around the first lens to back-reflect laser light and provide a reference to align the axis of the objective to the vacuum chamber viewports. For pulling out the spacing rings, notches have been milled in each of the rings. The technical drawings of the design can be obtained on request. We note that the moderate numerical aperture of the objective did not require extreme precautions concerning the deformations due to mounting. 

This simple stacking model meets the requirements stated above. The spacing rings can be milled with several tens of micrometer precision. This matches directly the requirement on the axial placement and, given the \SI{50.8}{\milli\meter} diameter of the rings, also matches the set requirement on the tilt of the lenses. The housing for our objective is made out of polyether ether ketone (PEEK), which is insulating and at the same time particularly convenient to machine. PEEK has a tensile modulus of \SIrange{3.1}{3.8}{\giga\pascal} and a linear coefficient of thermal expansion of \SIrange{4.7e-5}{5.5e-5}{\per\kelvin} \cite{2001VandeveldeKiekens}. We estimate that for \SI{1}{\kelvin} variation of the ambient temperature the thermal expansion of the supporting PEEK material will be just a few micrometers, and the lens will remain diffraction-limited. 

\section{Performance characterization}

\begin{figure}[t]
    \centering
    \begin{subfigure}[t]{\textwidth} 
     \centering
      \begin{tikzpicture}
      \begin{axis} [
          enlargelimits=false,
          axis equal image,
          axis on top,
          axis lines = none,
          xmin = 0, xmax = 1,
          ymin = 0, ymax = 1,
          width = \textwidth,
      ]
          \addplotgraphicsnatural[xmin = 0, xmax = 1,
          ymin = 0.15, ymax = 0.85] {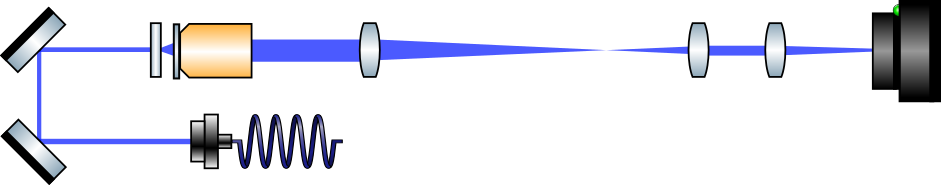};

          \draw[gray] (axis cs:0.25,0.15) -- (axis cs:0.25,0.18);
          \draw[gray] (axis cs:0.39,0.85) -- (axis cs:0.39,0.78);
          \draw[gray] (axis cs:0.1, 0.85) -- (axis cs:0.15,0.75);
          \draw[gray] (axis cs:0.185,0.85) -- (axis cs:0.185,0.78);
          \draw[gray] (axis cs:0.265,0.85) -- (axis cs:0.24,0.78);
          \draw[gray] (axis cs:0.82,0.85) -- (axis cs:0.82,0.78);
          \draw[gray] (axis cs:0.69,0.85) -- (axis cs:0.75,0.75);
          \draw[gray] (axis cs:0.9,0.15) -- (axis cs:0.95,0.45);

          \draw[color=black](axis cs:0.25,0.1) node[rotate=0] {Collimator};
          \draw[color=black](axis cs:0.40,0.9) node[rotate=0] {$f=\SI{500}{\milli\meter}$};
          \draw[color=black](axis cs:0.69,0.9) node[rotate=0] {$f=\SI{40}{\milli\meter}$};
          \draw[color=black](axis cs:0.85,0.9) node[rotate=0] {$f=\SI{125}{\milli\meter}$};
          \draw[color=black](axis cs:0.075,0.9) node[] {Target};
          \draw[color=black](axis cs:0.18,0.9) node[] {Viewp.};
          \draw[color=black](axis cs:0.265,0.9) node[] {Obj.};
          \draw[color=black](axis cs:0.9,0.1) node[] {CMOS Camera};

          \draw[black,->] (axis cs:0.65,0.25) -- (axis cs:0.6,0.25);
          \draw[black,->] (axis cs:0.65,0.25) -- (axis cs:0.65,0.43);
          \draw[color=black](axis cs:0.625,0.15) node[] {z};
          \draw[color=black](axis cs:0.675,0.325) node[] {y};

      \end{axis}
      \end{tikzpicture}
      \subcaption{}
      \label{fig:testsetupschamtic}
     \end{subfigure} \qquad \\

    \begin{subfigure}[t]{0.55\textwidth} 
        \begin{tikzpicture}
        \begin{axis} [
          baseline,
          xmin = 0, xmax = 1,
          ymin = 0, ymax = 1,
          xlabel=\phantom{$x [\si{\micro\meter}]$},
          ylabel=\phantom{$y [\si{\micro\meter}]$},
          axis lines=left, xticklabels=\empty, yticklabels=\empty,
          axis line style = {draw = none},
          tick style = {draw = none}
      ]
          \addplotgraphicsnatural[xmin = 0, xmax = 1,
          ymin = 0., ymax = 1] {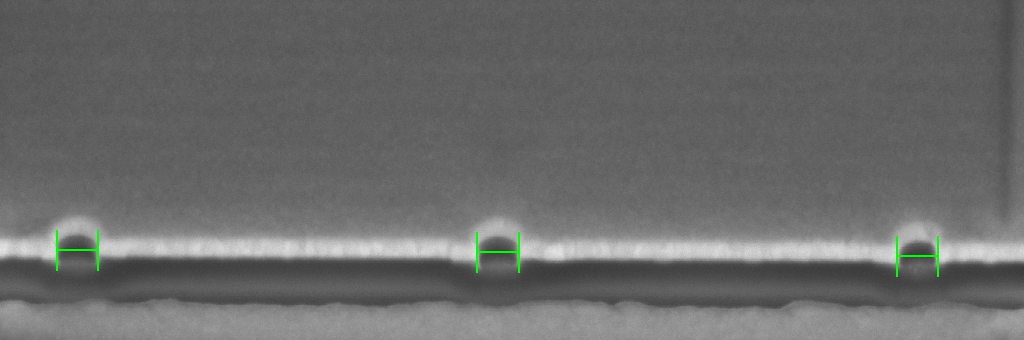};
          
          \draw[color=green](axis cs:0.12,0.6) node[] {\SI{198.3}{\nano\meter}};
          \draw[color=green](axis cs:0.5,0.6) node[] {\SI{203.2}{\nano\meter}};
          \draw[color=green](axis cs:0.88,0.6) node[] {\SI{198.3}{\nano\meter}};
        \end{axis}
        \end{tikzpicture}
        \subcaption{}
        \label{fig:field3spotssem}    
    \end{subfigure}
    \hfill
     \begin{subfigure}[t]{0.32\textwidth} 
    \centering
        \begin{tikzpicture}
        \begin{axis}[
        baseline,
        xlabel={$x [\si{\micro\meter}]$},
        ylabel={$y [\si{\micro\meter}]$},
        xlabel near ticks,
        ylabel near ticks,
        width = \textwidth,
        height = \textwidth,
        enlargelimits=false,
        xmin = 0.,
        xmax = 20.,
        ymin = 0.,
        ymax = 20.,
        grid = major
        ]
            \addplot+ [only marks] table {field3.dat};
        \end{axis}
        \end{tikzpicture}
        \subcaption{}
        \label{fig:field3spots}
    \end{subfigure}
    \quad
    \caption{Setup for characterizing the objective. (a) Overview of the setup. The target consists of a glass plate with a metal coating that has six \SI{200}{\nano\meter} diameter holes. The target, objective and the \SI{500}{\milli\meter} lens are mounted on translation stages. Drawing is not to scale. (b) Scanning electron microscope image of holes similar to those used in the target. (c) Schematic drawing of the locations of the nanoholes on the test target.  \label{fig:testsetup}}
\end{figure}
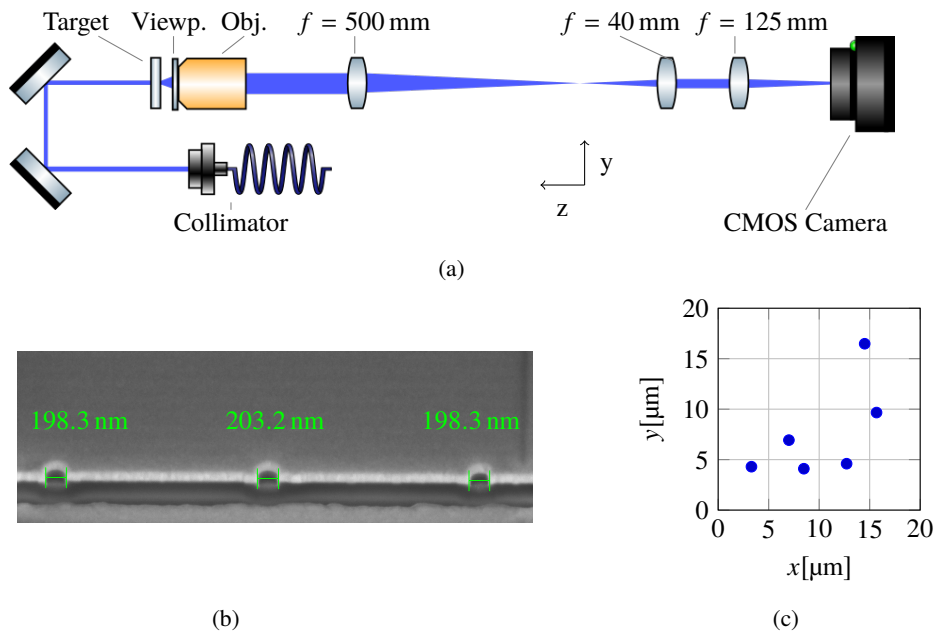

The resolution of the objective is obtained by directly measuring the point spread function (PSF) of the objective. A schematic of the setup used for testing the performance of the objective is shown in Fig. \ref{fig:testsetupschamtic}. In our case, the point source is satisfactorily substituted by a back-illuminated glass plate coated with a \SI{2}{\nano\meter} film of Cr and \SI{100}{\nano\meter} of Pt-Pd containing \SI{200}{\nano\meter} diameter holes provided by ETH Z\"urich. The residual transmission through the metallic coating itself is negligible for all the wavelengths used in our work and complex light propagation effects through such short tunnels are not considered here \cite{Popov2007JOSAA, Abajo2002OE}. A scanning electron microscope image of holes as the ones used in the target is shown in Fig. \ref{fig:field3spotssem}. Using the Rayleigh criterion, the resolution for a point light source is defined as the distance of the principle maximum of the PSF to the first minimum \cite{Rayleigh1879, Saleh2013FoP}.  For our objective with a predicted resolution of \SI{639}{\nano\meter} the \SI{200}{\nano\meter} diameter holes can be considered point sources (the PSF approximated by a Gaussian changes by less than \SI{10}{\nano\meter} if it is convoluted with the finite size holes). 
Nanoholes have also been used to measure PSFs in \cite{Tapashree2014PSF}. The positioning of the holes is chosen to be non-periodic to avoid artifacts stemming from the Talbot effect \cite{Talbot1836, Saleh2013FoP} and is shown in Fig. \ref{fig:field3spots}. The objective images the diffraction pattern from the illuminated holes using a $f=\SI{500}{\milli\meter}$ achromatic doublet as field lens. To increase the magnification, $f=\SI{40}{\milli\meter}$ and $f=\SI{125}{\milli\meter}$ achromatic doublets are placed in a telescope configuration. The extra telescope gives us the flexibility to control the beam size at the sensor and avoid limitations coming from the pixel size. The magnified pattern is imaged using a BlackFly complementary metal-oxide semiconductor (CMOS) camera with pixel size \SI{5.6}{\micro\meter}. For ease of alignment, the target (x, y-direction), objective (z-direction) and field lens (z-direction) are placed on linear translation stages. 

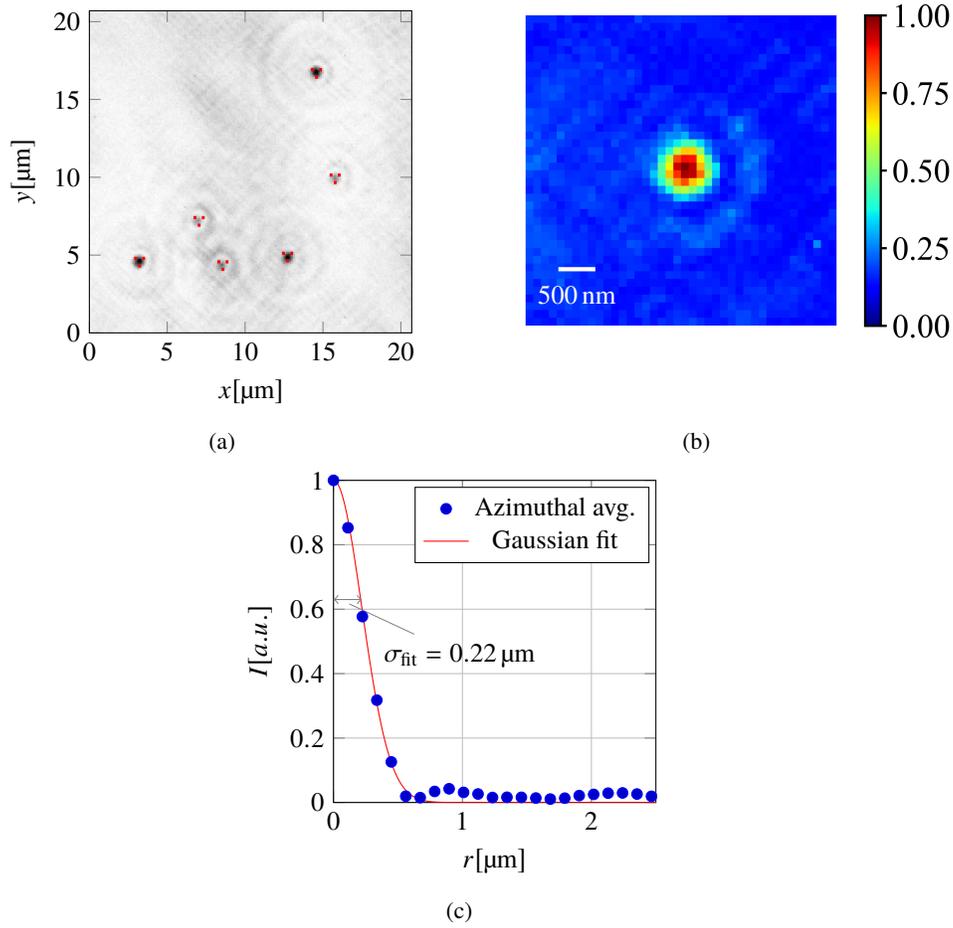
\begin{figure}[t]
    \hspace{-1cm}
    \begin{subfigure}[t]{0.44\textwidth}
        \begin{tikzpicture}
        \begin{axis}[
            baseline,
            enlargelimits=false,
            axis on top,
            width = \textwidth,
            height = \textwidth,
            baseline,
            xlabel = {$x [\si{\micro\meter}]$},
            ylabel = {$y [\si{\micro\meter}]$},
            ylabel near ticks,
            xlabel near ticks
        ]
            \addplot graphics [
                xmin=0,xmax=20.7,
                ymin=0,ymax=20.7]
                {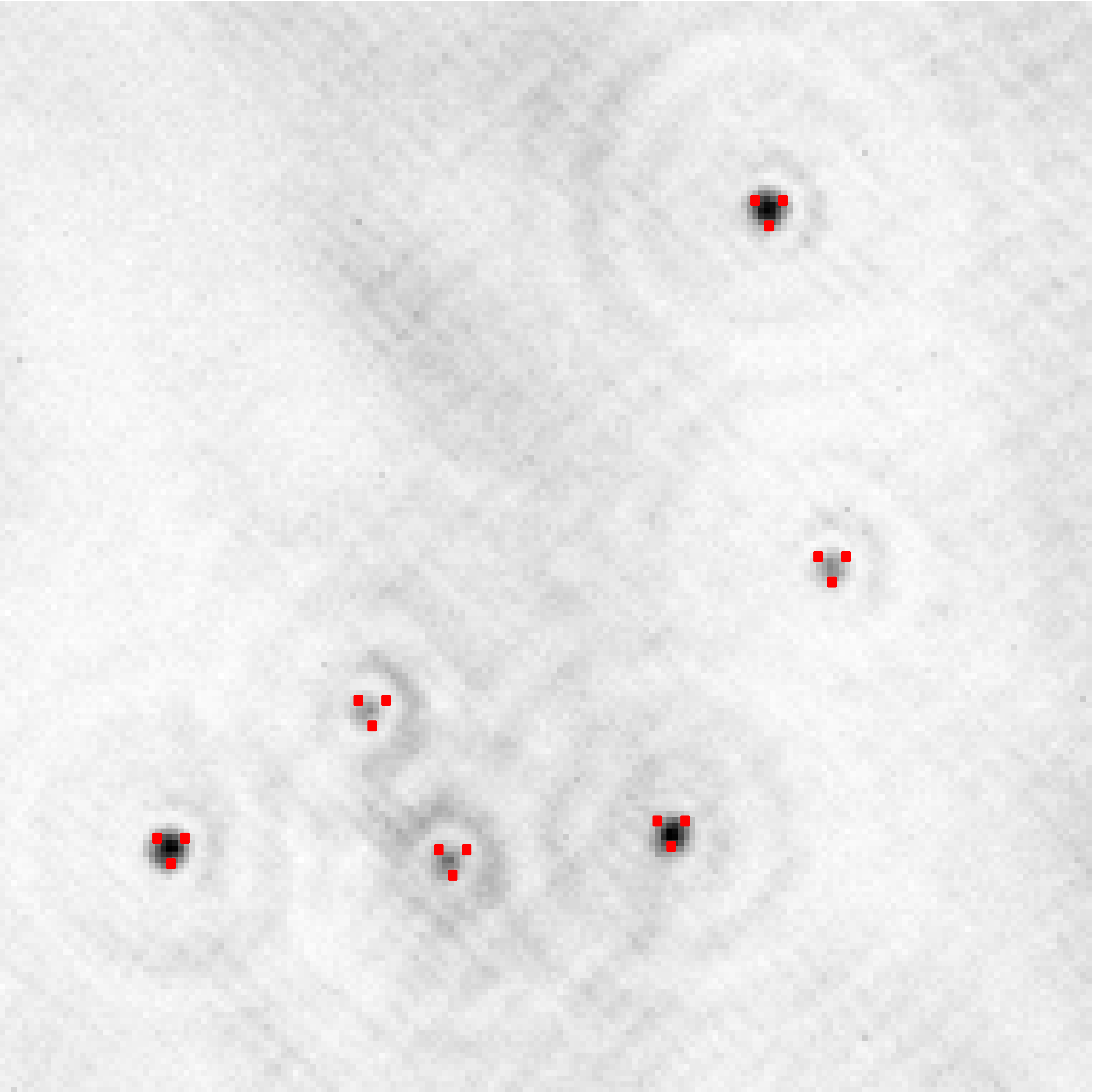};
        \end{axis}
        \end{tikzpicture}
        \caption{}
        \label{fig:psftestsetupcrosshairs}
    \end{subfigure}
    \quad
     \begin{subfigure}[t]{0.44\textwidth}
        \begin{tikzpicture}[] 
        \begin{axis}[white][
        \pgfplotsset{baseline,xtick style= {draw=none},ytick style= {draw=none},yticklabels={,,}, hide y axis}];

        \addplotgraphicsnatural[xmin = 0, xmax = 2,
          ymin = 0, ymax = 2] {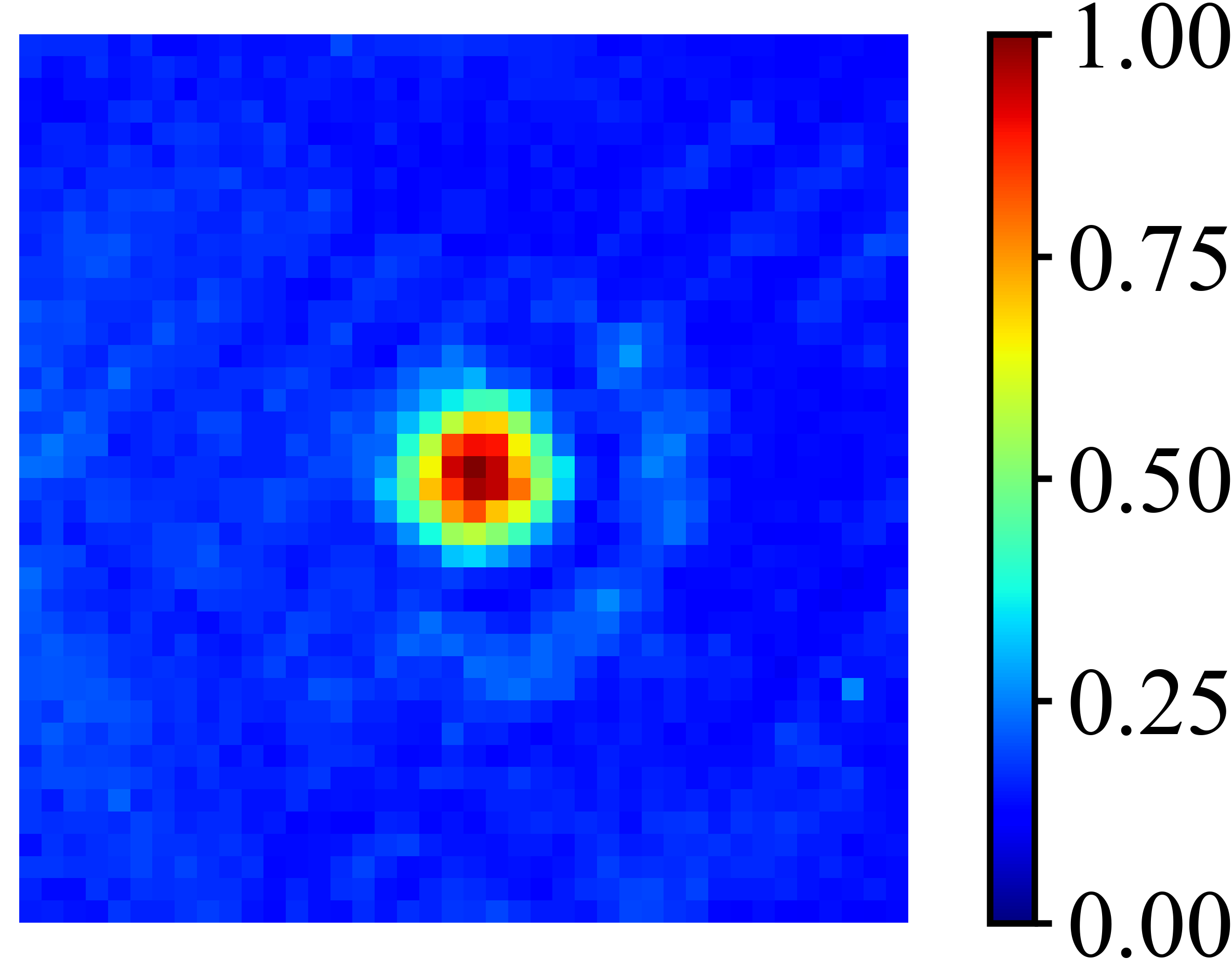};
           
                \draw [white, ultra thick,-] (axis cs:0.18,0.43) -- (axis cs:0.352,0.43);
                \node at (axis cs:0.266,0.28) {$\SI{500}{\nano\meter}$};
        \end{axis}
        \end{tikzpicture} 
        \caption{}
        \label{fig:PSFisolated}
    \end{subfigure} \\
    \hspace{-1cm}
    \begin{subfigure}[t]{0.44\textwidth} 
        \begin{tikzpicture}
        \begin{axis}[
        baseline,
        xlabel={$r [\si{\micro\meter}]$},
        ylabel={$I [a.u.]$},
        width = 1\textwidth,
        height = 1\textwidth,
        enlargelimits=false,
        xmin = 0.,
        xmax = 2.5,
        ymin = 0.,
        ymax = 1.,
        xlabel near ticks,
        ylabel near ticks,
        grid = major,
        legend entries = {Azimuthal avg.,Gaussian fit}
        ]
            \addplot+ [only marks] table {removebackgroundrun4.dat};
            \addplot [red] table {removebackgroundrun4fitgaussian.dat};
            \draw [gray, <->] (axis cs:0,0.63) -- (axis cs:0.22,0.63);
            \node[color=black,pin=315:{$\sigma_{\text{fit}} = \SI{0.22}{\micro\meter}$}] at (axis cs:0.05,0.63) {};
        \end{axis}
        \end{tikzpicture}
        \caption{}
        \label{fig:psftestsetupfit}
    \end{subfigure}
    
    \caption{Analysis of the objective's resolution. (a) Using a Laplacian of Gaussian algorithm each spot center is detected and marked at the center of the three red dots. (b) A typical image of the observed PSF without background subtraction. (c) An azimuthal average of the intensity of a single point. The red line shows a Gaussian fit through the data resulting in $\sigma_{\text{fit}} = \SI{0.22}{\micro\meter}$, which corresponds to a resolution of $\SI{0.639}{\micro\meter}$ for \SI{461}{\nano\meter} light.}
    \label{fig:analysispsftestsetup}
\end{figure}

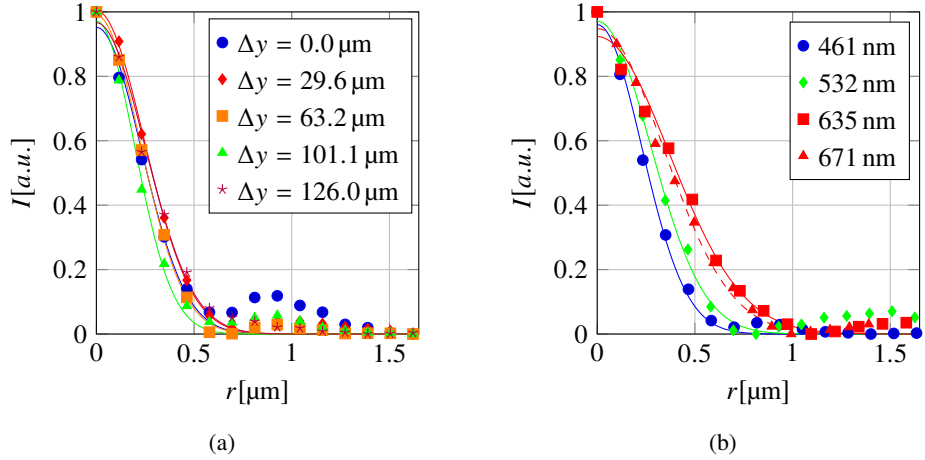
\begin{figure}
    \centering
    \begin{subfigure}[t]{0.44\textwidth}
        \begin{tikzpicture}
        \begin{axis}[
        baseline,
        xlabel={$r [\si{\micro\meter}]$},
        ylabel={$I [a.u.]$},
        xlabel near ticks,
        ylabel near ticks,
        width = \textwidth,
        height = \textwidth,
        enlargelimits=false,
        xmin = 0.,
        xmax = 1.65,
        ymin = 0.,
        ymax = 1.,
        grid = major,
        legend entries = {$\begin{aligned}[t]&\Delta y = \SI{0.0}{\micro\meter}\\[-1pt]&\Delta y = \SI{29.6}{\micro\meter}\\[-1pt]&\Delta y = \SI{63.2}{\micro\meter}\\[-1pt]&\Delta y = \SI{101.1}{\micro\meter}\\[-1pt]&\Delta y = \SI{126.0}{\micro\meter}\end{aligned}$,\strut,\strut,\strut,\strut}            
    ]
            \addplot+ [blue,only marks,mark color=blue] table {usethisfov1.dat};
            \addplot+ [red,only marks,color=red,mark=diamond*,mark options={solid,fill=red}] table {usethisfov2.dat};
            \addplot+ [only marks,color=orange,mark=square*,mark options={solid,fill=orange}] table {usethisfov3.dat};
            \addplot+ [only marks,color=green,mark=triangle*,mark options={solid,fill=green}] table {usethisfov4.dat};
            \addplot+ [only marks,color=purple,mark=star,mark options={solid,fill=purple}] table {usethisfov5.dat};
            \addplot [blue] table {usethisfov1fit.dat};
            \addplot [red] table {usethisfov2fit.dat};
            \addplot [orange] table {usethisfov3fit.dat};
            \addplot [green] table {usethisfov4fit.dat};
            \addplot [purple] table {usethisfov5fit.dat};
                
        \end{axis}
        \end{tikzpicture}
        \caption{}
        \label{fig:fovradialprofiles}
    \end{subfigure}
    \qquad
    \begin{subfigure}[t]{0.44\textwidth} 
            \begin{tikzpicture}
            \begin{axis}[
            baseline,
            xlabel={$r [\si{\micro\meter}]$},
            ylabel={$I [a.u.]$},
            xlabel near ticks,
            ylabel near ticks,
            width = \textwidth,
            height = \textwidth,
            enlargelimits=false,
            xmin = 0.,
            xmax = 1.65,
            ymin = 0.,
            ymax = 1.,
            grid = major,
            legend cell align=left,
            legend style={legend pos=north east},
            legend entries = {$\begin{aligned}[t]\SI{461}{\nano\meter}\\[-1pt] \SI{532}{\nano\meter}\\[-1pt] \SI{635}{\nano\meter}\\[-1pt]
            \SI{671}{\nano\meter}\\\end{aligned}$,\strut,\strut,\strut}
            ]
                \addplot+ [blue,only marks,mark color=blue] table {usethis461.dat};
                \addplot+ [green,only marks,color=green,mark=diamond*,mark options={solid,fill=green}] table {532new.dat};
                \addplot+ [red,only marks,color=red,mark=square*,mark options={solid,fill=red}] table {usethis635.dat};
                \addplot+ [red,only marks,color=red,mark=triangle*,mark options={solid,fill=red}] table {usethis671.dat};
                \addplot [blue] table {461fit.dat};
                \addplot [green] table {532newfit.dat};
                \addplot [red] table {635fit.dat};
                \addplot [red,dashed] table {671fit.dat};
            \end{axis}
            \end{tikzpicture}
            \caption{}
            \label{fig:differentlambda}
        \end{subfigure}
        \caption{Radial profiles of target hole images in dependence of (a) displacement $\Delta y$ of the target hole perpendicularly to the optical axis (for the determination of the FOV) or (b) wavelength. Solid lines depict Gaussian fits to the data.}
\end{figure}

In Fig. \ref{fig:psftestsetupcrosshairs}, the observed pattern from the illuminated nanoholes is shown. The radial symmetry of the PSFs indicates that aberrations are predominantly spherical. The three red dots denote the positions of the centers found using a Laplacian of Gaussian algorithm. We perform a calibration of the image by calculating the ratio between the distance in pixels and the scanning electron microscope measured distance in \si{\micro\metre} for each pair of holes. In the case of Fig. \ref{fig:psftestsetupcrosshairs}, this procedure gives an average calibration of \SI{9.23(6)}{px\per\micro\meter} over 15 distances. The calibration was done at several distances from the sample without any influence on the magnification due to imperfect alignment with respect to the optical axis. A typical image of the PSF close to the apparent focus without any background subtraction is shown in Fig. \ref{fig:PSFisolated}. At each of the spot centers an azimuthal averaging is performed to obtain a radial profile of the PSF of the objective for that spot, to which a Gaussian fit is made for the resolution, given by
$ \label{eqn:gaussianfitfunc} I = I_0 \exp{\left(-r^2/(2\sigma^2)\right)} + a, $ where $I_0$ is the peak intensity, $\sigma$ is the standard deviation of the Gaussian and $a$ is a constant for the read-out noise. The radial profile and fit for one nanohole is presented in Fig. \ref{fig:psftestsetupfit}. Here background noise is taken to be the azimuthal average far from the spot and is subtracted. The resolution of the objective is found as the average of the six fits (corresponding to the six nanoholes) and the error reported here is purely statistical and is defined by the standard deviation in the sample, yielding \SI{0.63(4)}{\micro\meter}. We note that for nanoholes separated by distances well within the FOV the systematic error introduced by their lateral distance can be neglected as shown in Fig. \ref{fig:fovradialprofiles}.

\begin{figure}
\begin{subfigure}{0.95\textwidth}
  \centering
        \begin{tikzpicture}
        \begin{axis}[
        baseline,
        xlabel={$z [\si{\micro\meter}]$},
        ylabel={$r [\si{\micro\meter}]$},
        width = 0.8\textwidth,
        height = 0.3\textwidth,
        enlargelimits=false,
        xmin = -3.0,
        xmax = 0.5,
        ymin = 0.,
        ymax = 1.0,
        xlabel near ticks,
        ylabel near ticks,
        ytick={0,0.5,1}
        ]
        \addplot graphics [xmin = -3.5, xmax = 0.5,
          ymin = 0, ymax = 1.0] {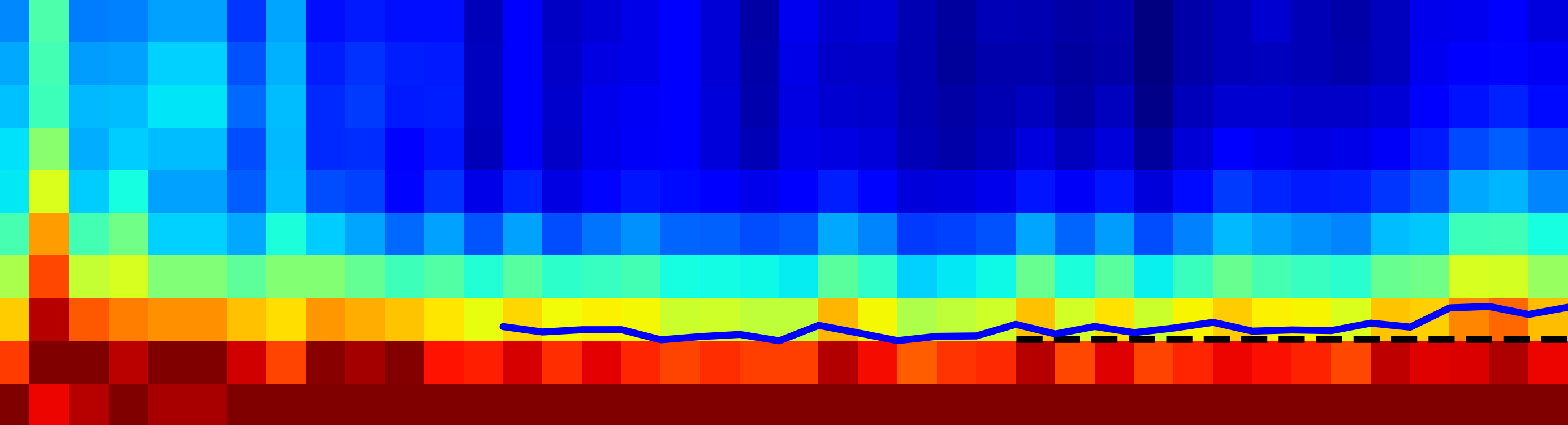};
        \end{axis}
        \end{tikzpicture}
  \caption{}
  \label{fig:3dpsfa}
\end{subfigure}
\begin{subfigure}{0.95\textwidth}
  \centering
        \begin{tikzpicture}
        \begin{axis}[
        baseline,
        xlabel={$z [\si{\micro\meter}]$},
        ylabel={$r [\si{\micro\meter}]$},
        width = 0.8\textwidth,
        height = 0.3\textwidth,
        enlargelimits=false,
        xmin = -3.0,
        xmax = 0.5,
        ymin = 0.,
        ymax = 1.,
        xlabel near ticks,
        ylabel near ticks,
        ytick={0,0.5,1}
        ]
        \addplot graphics [xmin = -3.5, xmax = 0.5,
          ymin = 0, ymax = 1] {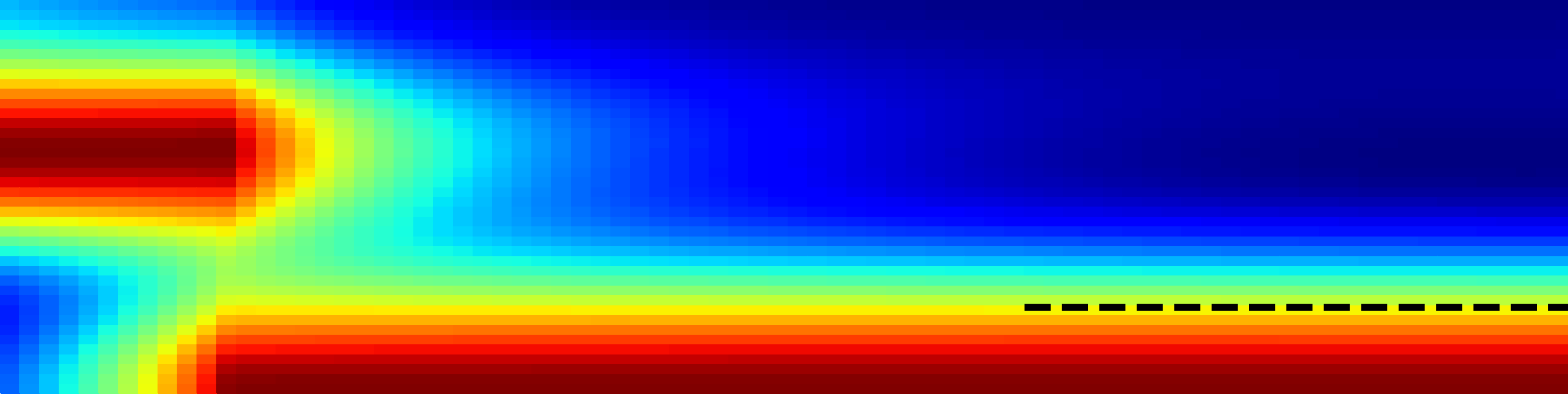};
        \end{axis}
        \end{tikzpicture}
  \caption{}
  \label{fig:3dpsfb}
\end{subfigure}
\begin{subfigure}{0.95\textwidth}
  \centering
        \begin{tikzpicture}
        \begin{axis}[
        baseline,
        xlabel={$z [\si{\micro\meter}]$},
        ylabel={$r [\si{\micro\meter}]$},
        width = 0.8\textwidth,
        height = 0.3\textwidth,
        enlargelimits=false,
        xmin = -3.0,
        xmax = 0.5,
        ymin = 0.,
        ymax = 1.,
        xlabel near ticks,
        ylabel near ticks,
        ytick={0,0.5,1}
        ]
        \addplot graphics [xmin = -3.5, xmax = 0.5,
          ymin = 0, ymax = 1] {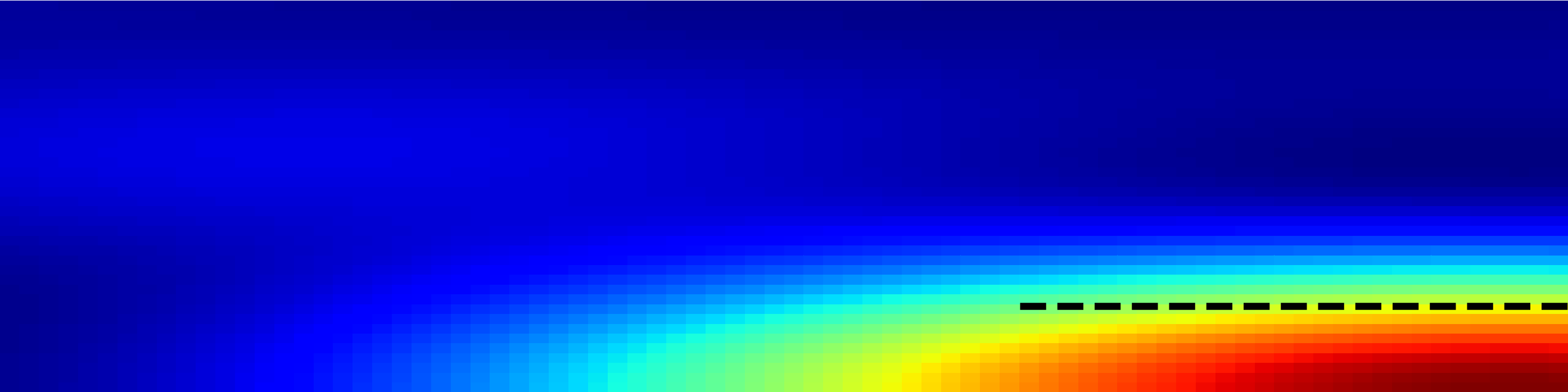};
        \end{axis}
        \end{tikzpicture} 
  \caption{}
  \label{fig:3dpsfc}
\end{subfigure}
\caption{Through-focus PSF of the objective. The radial profile of the PSF is plotted for a range of $z$-values around the focus. (a) Experimental data with normalization to the maximum of intensity at each location. The blue solid line denotes the standard deviation, $\sigma_{\text{fit}}$, for each Gaussian fit. The black dashed line denotes the diffraction limit for \SI{461}{\nano\meter} light of \SI{639}{\nano\meter} ($ \sigma_{\text{fit}} =$ \SI{0.22}{\micro \meter}). (b) The corresponding radial profile simulated using the Extended Nijboer-Zernike diffraction theory with the same normalization as in (a). (c) The same as (b) but with the values normalized to the global maximum of the PSF. For (b) and (c) the black dashed line stops at the point where the theoretical Strehl ratio drops below 0.8. The colormap is identical with the one in Fig. \ref{fig:analysispsftestsetup}. The objective is predicted to be diffraction-limited from around \SIrange{-1.0}{1.3}{\micro\meter}.}
\label{fig:3dpsf}
\end{figure}
The DOF that corresponds to our numerical aperture for focusing in vacuum $\left(n=1\right)$ is $\lambda \sqrt{n^2-N\!A^2}/N\!A^2=\SI{2.1}{\micro\meter}$ \cite{Shillaber1944DOF}. The PSF as a function of position along the optical axis is measured by shifting the lens. Fig. \ref{fig:3dpsfa} shows the experimentally obtained normalized radial profile of the objective along this axis. Since for not properly focused spots, the Laplacian of Gaussian algorithm does not return satisfactory results, a single spot is cropped out and a 2D Gaussian fit is used to determine the center of the spot. For z-values far from the focus (approximately beyond $\SI{-2.5}{\micro\meter}$ in Fig. \ref{fig:3dpsfa}), the central cannot be fitted properly and the resulting radial profiles are not a reliable measure. The blue line in Fig. \ref{fig:3dpsfa} denotes the values of the fitted resolution for the radial profile and the black dashed line lies at the diffraction limit of \SI{639}{\nano\meter}. We mark distance 0 in Fig. \ref{fig:3dpsfa} based on qualitative matching with Fig. \ref{fig:3dpsfb} for which distance 0 was determined from our simulations.  The objective based on Gaussian fits leads to resolutions close to the the diffraction limit in a range from \SIrange{-2.2}{0.2}{\micro\meter}. That is close to the predictions from our simulations that use the stricter criterion of the Strehl ratio being larger than 0.8 and the encircled energy in the first lobe exceeding 68$\%$ \cite{Guenther2015MO}. The evolution of the PSF expansion with axial distance is simulated using the extended Nijboer-Zernike (ENZ) diffraction theory \cite{Janssen2002aENZ, Braat2002ENZ ,Janssen2008ENZ, VanHaver2010PhD} taking as input the Zernike polynomial aberration coefficients obtained from our design software. The contributions on the RMS wavefront distortion of the low-order Zernike polynomials in $\lambda$(\SI{460.733}{\nano\meter}) units are: defocus $\left(2,0\right)=-0.0497$, spherical $\left(4,0\right)=-0.0167$, astigmatism $\left(2,2\right)=-0.000377$, coma $\left(3,1\right)=-0.000289 $, and also secondary spherical $\left(6,0\right)=0.046135$. The qualitative agreement between our zero free parameter model and the experimental data is reasonable. From these values it is evident that defocus and secondary spherical aberrations are the main contributions to the wavefront distortion. To verify the diffraction-limited performance of our objective around our main imaging wavelength we have also numerically obtained a high on-axis Strehl ratio of 0.95 that is larger than the required 0.8 ratio for a range of \SI{0.3}{\nano\meter} without any refocusing. When the Strehl ratio drops below 0.8 the calculated RMS radius increases from \SI{363}{\nano\meter} to \SI{432}{\nano\meter}. Our simulations for the wavelengths of \SI{461}{\nano\meter}, \SI{532}{\nano\meter}, and \SI{689}{\nano\meter} give also a chromatic focus shift (positive values indicate shifts away from the objective) of \SI{3.7}{\micro\meter}/\SI{}{\nano\meter}.

A FOV that can permit the imaging and manipulation of hundreds of particles held in an optical lattice is considered large enough for our purposes. We measure the FOV by translating the target in the $x, y$-plane.  Azimuthally integrated radial profiles for six positions during a translation of $\Delta y$ along $y$ are presented in Fig. \ref{fig:fovradialprofiles}. All the profiles have been normalized to the maximum of intensity and the averaged background far from the nanoholes is subtracted. The azimuthal average is justified given that the dominant aberrations are found to be the spherical ones. Since several tens of pixels are required for a well-resolved PSF the total number of pixels on the chip limit the measurement range to $\Delta y= \SI{126}{\micro\meter}$. The objective has a resolution of \SI{0.63(4)}{\micro\meter} for a displacement of up to \SI{100}{\micro\meter}, which is smaller than the predicted FOV. The limited size of the CCD chip, together with the subjective placement of the chip by eye, might lead to such an underestimation. 

For manipulation of strontium atoms for quantum simulation \SI{671}{\nano\meter} light can be used because it addresses an ultranarrow transition. Using the same setup, we have tested the objective for different wavelengths. The resulting radial profiles (with normalization to their maximum intensity and background subtraction) after refocusing are presented in Fig. \ref{fig:differentlambda} for \SI{461}{\nano\meter}, \SI{532}{\nano\meter}, \SI{635}{\nano\meter} and \SI{671}{\nano\meter} light. The measured resolutions (in parentheses: the diffraction limit)  are \SI{0.63(4)}{\micro\meter} (\SI{0.639}{\micro\meter}), \SI{0.75(4)}{\micro\meter} (\SI{0.738}{\micro\meter}), \SI{1.09(9)}{\micro\meter} (\SI{0.880}{\micro\meter}) and \SI{1.05(4)}{\micro\meter} (\SI{0.930}{\micro\meter}), respectively. The measured resolutions for both \SI{461}{\nano\meter} and \SI{532}{\nano\meter} are close to the diffraction limit within the margin. The resolution is close to \SI{1}{\micro\meter} for \SI{671}{\nano\meter} light. With respect to some state-of-the-art representative microscopes capable of imaging quantum gases such as \cite{Bakr2010QGM, Cheuk2015FQGM, Alberti2016MO, Robens2017MO} our work achieves moderate numerical aperture, but because of the short wavelength of the atomic transition the resolution is comparable. At the same time, the requirement for placing optics inside vacuum is lifted due to the long working distance of our objective. Commercial lenses have long working distances similar to the ones that we demonstrate here but the available sizes would not leave enough space for our magnetic coils.  

The most relevant technical characteristics of our objective are summarized in Table \ref{tab:lenssummary}.

\begin{table}[t]

  \renewcommand{\arraystretch}{1.2}
    \centering
     
    \begin{tabular}{c c}
        \hline \hline
         Resolution & \SI{0.63(4)}{\micro\meter}\\
         Numerical aperture & 0.44\\
         Strehl ratio (on-axis) & 0.95\\
         Chromatic bandwidth (without refocusing)  &\SI{0.3}{\nano\meter} \\  
         Chromatic focus shift (estimation) &\SI{3.7}{\micro\meter}/\SI{}{\nano\meter} \\  
         Total collection angle & $2 \sin^{-1}\left(\text{NA}\right) = \SI{0.91}{\radian}$\\
         Working distance & \SI{18}{\milli\meter}\\
         Field of view & $\geq \SI{100}{\micro\meter}$\\
         Effective focal length & \SI{25}{\milli\meter}\\
         Depth of field & $\lambda \sqrt{n^2-N\!A^2}/N\!A^2=\SI{2.1}{\micro\meter}$\\
        \hline \hline
    \end{tabular}
  \caption{Technical characteristics of the microscope objective  at \SI{461}{\nano\meter}.}\label{tab:lenssummary}
\end{table}

\section{Conclusions}
We have presented a design for a microscope objective consisting only of commercially available lenses that is diffraction-limited for \SI{461}{\nano\meter} light. The objective is built using a simple stacking design and is measured to have a resolution of \SI{0.63(4)}{\micro\meter}, which is in agreement with the predicted value. This performance, together with the near-diffraction limited performance for \SI{532}{\nano\meter} light makes this design useful for both quantum gas microscopes and optical tweezer experiments with strontium. It also makes this design an interesting candidate for experiments with other atomic species such as erbium (\SI{401}{\nano\meter}), ytterbium (\SI{399}{\nano\meter}), and dysprosium (\SI{421}{\nano\meter}), as well as Rydberg experiments with rubidium (\SI{420}{\nano\meter} and \SI{480}{\nano\meter}).

\section*{Funding}

This project has received funding from the European Research Council (ERC) under the European Union's Seventh Framework Programme (FP7/2007-2013) (Grant agreement No. 615117 QuantStro). We thank the Netherlands Organisation for Scientific Research (NWO) for funding through Vici grant No. 680-47-619 and Gravitation grant No. 024.003.037, Quantum Software Consortium. G.S. thanks the European Commission for Marie Sk\l{}odowska-Curie grant SYMULGAS, No. 661171. 

\section*{Acknowledgments}

We thank Hans Ellermeijer from the Unversiteit van Amsterdam Technology Center for his assistance in engineering the microscope objective.  We also thank Joakim Reuteler of ScopeM, ETH Z\"urich for the manufacturing of the test target. 

\section*{Disclosures}

The authors declare no conflicts of interest. Technical drawings of our objective can be provided upon reasonable request. 

\bibliography{main}

\pagebreak

\end{document}